\documentclass[format=sigconf]{acmart}

\usepackage{booktabs} 
\usepackage{colortbl}
\usepackage{graphicx}

\usepackage[normalem]{ulem}
\usepackage[inline]{enumitem}
\usepackage{multirow}
\usepackage{algorithm}
\usepackage[noend]{algpseudocode}

\algnewcommand\algorithmicforeach{\textbf{for each}}
\algdef{S}[FOR]{ForEach}[1]{\algorithmicforeach\ #1\ \algorithmicdo}

\setcitestyle{square}

\begin{document}

\title[RLFC]{RLFC: Random Access Light Field Compression using Key Views and Bounded Integer Sequence Encoding}

\author{Srihari Pratapa}
\affiliation{%
 \department{Department of Computer Science }
 \institution{University of North Carolina at Chapel Hill}}
\email{psrihariv@cs.unc.edu}

\author{Dinesh Manocha}
\affiliation{%
\department{ Department of Computer Science and Electrical \& Computer Engineering }
 \institution{University of Maryland at College Park}}
\email{dm@cs.umd.edu}




\begin{abstract}

We present a new hierarchical compression scheme for encoding light field images (LFI) that is suitable for interactive rendering. Our method (RLFC) exploits redundancies in the light field images by constructing a tree structure. The top level (root) of the tree captures  the common high-level details across the LFI, and  other levels (children) of the tree capture specific low-level details of the LFI.  Our decompressing algorithm corresponds to tree traversal operations and gathers the values stored at different levels of the tree. Furthermore, we use bounded integer sequence encoding which provides random access and fast hardware decoding for compressing the blocks of children of the tree. We have evaluated our method for 4D two-plane parameterized light fields. The compression rates vary from $0.08 - 2.5$ bits per pixel (bpp), resulting in compression ratios of around 200:1 to 20:1 for a PSNR quality of 40 to 50 dB. The decompression times for decoding the blocks of LFI are $1 - 3$ microseconds per channel on an NVIDIA GTX-960 and we can render new views with a resolution of $512\times512$ at $200$ fps. Our overall scheme is simple to implement and involves only bit manipulations and integer arithmetic operations.\footnote{website: \url{http://gamma.cs.unc.edu/LFC/}}

\end{abstract}


\copyrightyear{2019}
\acmYear{2019}
\setcopyright{acmcopyright}
\acmConference[I3D '19]{Symposium on Interactive 3D Graphics and Games}{May 21--23, 2019}{Montreal, QC, Canada}
\acmBooktitle{Symposium on Interactive 3D Graphics and Games (I3D '19), May 21--23, 2019, Montreal, QC, Canada}
\acmPrice{15.00}
\acmDOI{10.1145/3306131.3317018}
\acmISBN{978-1-4503-6310-5/19/05}


\begin{teaserfigure}
  \centering
  \includegraphics[width=\textwidth, keepaspectratio=true]{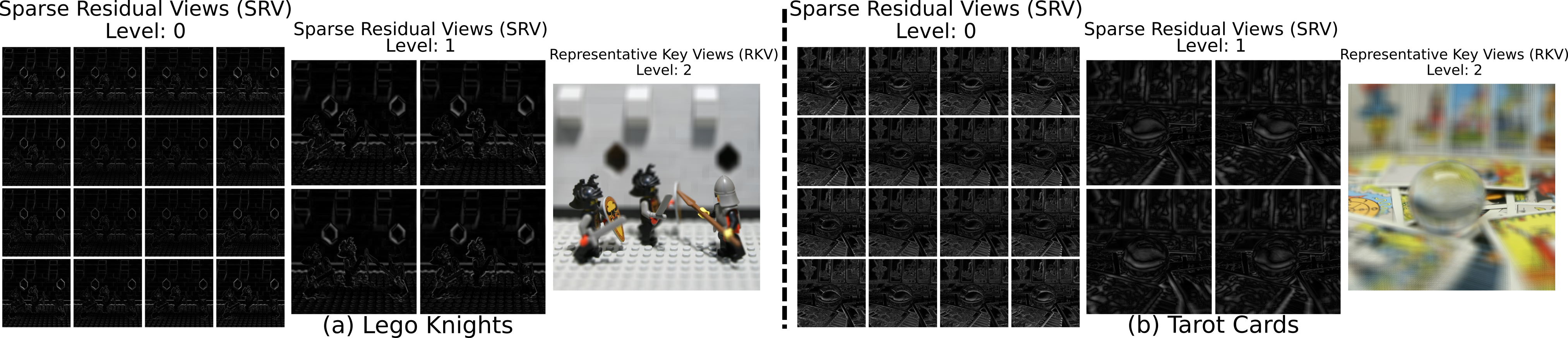}
  \caption{Our method (RLFC) is based on computing a hierarchy of new images: Representative Key
  Views (RKV) which capture the redundancies of the LF in the top levels of the hierarchy and
  Sparse Residual Views (SRV) that capture specific details of the LF in the bottom levels of the
  hierarchy. We present a visualization of the images (RKV \& SRV) computed in RLFC hierarchy on a
  small sample grid of the LF for two datasets. RLFC achieves compression by removing the
  insignificant details and exploiting the sparsity of the SRV images in the hierarchy.
  }
\end{teaserfigure}

\maketitle

\section{Introduction}
\label{sec:intro}



Light fields create a photo-realistic rendering of extremely complex scenes that are difficult to achieve with other conventional rendering techniques. The photo-realistic rendering from a light field (LF) makes virtual reality (VR) content more immersive and improves the sense of presence in real world-scenes. Levoy and Hanrahan~\citeyear{LFLevoy96} and Gortler et al.~\citeyear{LFGortler96} describe LF rendering methods by capturing a scene using a camera grid constructed using camera arrays. This has been an active area of research for more than two decades and many improved methods for capturing and rendering have been proposed.
Several hand held plenoptic cameras have been developed to capture the high stereo light fields of real scenes~\cite{CAMng2005light,CAMperwass2012single} and to use for VR applications ~\cite{LFYu2017}.

One of the primary challenges in using image-based rendering (IBR) approaches based on LFI is the amount of data needed to capture the 3D scenes. Such IBR techniques generate a lot of images to sample the light rays for a given scene. Typically, the size of the uncompressed light field images can vary from $200$MB to $10$GB (or more) and can be even larger depending on the sampling rate and image resolution. Several methods and hardware techniques for interactive rendering of light fields have been proposed~\cite{LFInteChen2002,LFInterjones2007rendering}. The need to use high resolution light fields has increased with recent demand for high resolution multimedia content ($2K$ or $4K$ resolution). An efficient way to capture high resolution panoramic light fields is discussed in ~\citet{LFInteGigabirklbauer2013rendering}.


In order to store, transmit, and render LFI, it is important to develop good compression algorithms. Different schemes have been proposed for compressing LFI, as surveyed in ~\cite{LFSurvey8010398}. The majority of these methods provide high compression rates, like standard 2D image compression methods, but they require decoding all the LFI samples into memory before rendering. During rendering, any of the sampled light rays (pixels) from the LFI  may be used for computing new views. For interactive rendering in VR and mobile devices, it is necessary for the compressed LFI bitstream to have random access capabilities. Random access to the compressed LFI bitstream can reduce the memory footprint during rendering by a significant factor. 


\textbf{Main Results:}  
We present a new hierarchical compression scheme (RLFC) for encoding light field images. The primary application of our method is for interactive rendering in VR and mobile devices, which require low latency and low memory footprint. Our method is based on clustering spatially close sampled images of light fields and constructing a tree. After the construction process is done, the root of the tree stores the common features or characteristics among the LFI; the rest of the children nodes of the tree store the specific low-level or high-frequency details of the LFI. In our tree construction process we compute two types of new images; images that capture the common details (redundancies) among the LFI are referred to as \textit{representative key views (RKVs)} and the images that capture high-frequency details are referred to as \textit{sparse residual views (SRVs)}. 
We start the construction of the tree in a bottom-up manner computing multiple levels of RKVs and at the start of the building process the bottom level of the tree is initialized with the original LFI. 
Next, we proceed in a top-down manner and compute \textit{SRVs} between alternate levels of \textit{RKVs} in the tree. 
Once the tree is computed, only the top-level \textit{RKVs} and the \textit{SRVs} of the tree are stored. 

The top-level \textit{RKVs} are compressed using standard image compression techniques (e.g., JPEG2000). The SRVs correspond to low-level details of the LFI and they are sparse in terms of the features captured. At all levels, the SRVs are divided into blocks and only the blocks with significant details are stored. Moreover,  significant blocks of the SRVs at each level of the tree are encoded using Bounded Sequence Integer Encoding (BISE)~\cite{ASTC}. The resulting compressed bitstream is represented to support selective decoding and provides random access to blocks of pixels with only one level of indirection.
We have evaluated our method on the two-plane parameterization of light fields and present the results on the Stanford light field archive. The decoding time to decompress a block of pixels from our compressed stream is 1-3 microseconds on an NVIDIA GTX-960 GPU and RLFC can be used to render new views with a resolution of $512\times512$ at $200$ fps. We obtain compression ratios of around $200:1$ to $20:1$ for a PSNR quality of $40-50$ dB.

The remainder of the paper is organized as follows: Section~\ref{sec:back} gives an overview of prior work in light field rendering and compression. Section~\ref{sec:method} gives  details about compression, decompression, and interactive rendering. In Section~\ref{sec:results}, we present a detailed analysis of our compression scheme and highlight the results on various benchmarks.

\section{Background}
\label{sec:back}
In this section, we present a brief overview of light field rendering and compression algorithms.

  \begin{figure*}[t!]
 \centering
 \includegraphics[width=\textwidth, keepaspectratio=true]{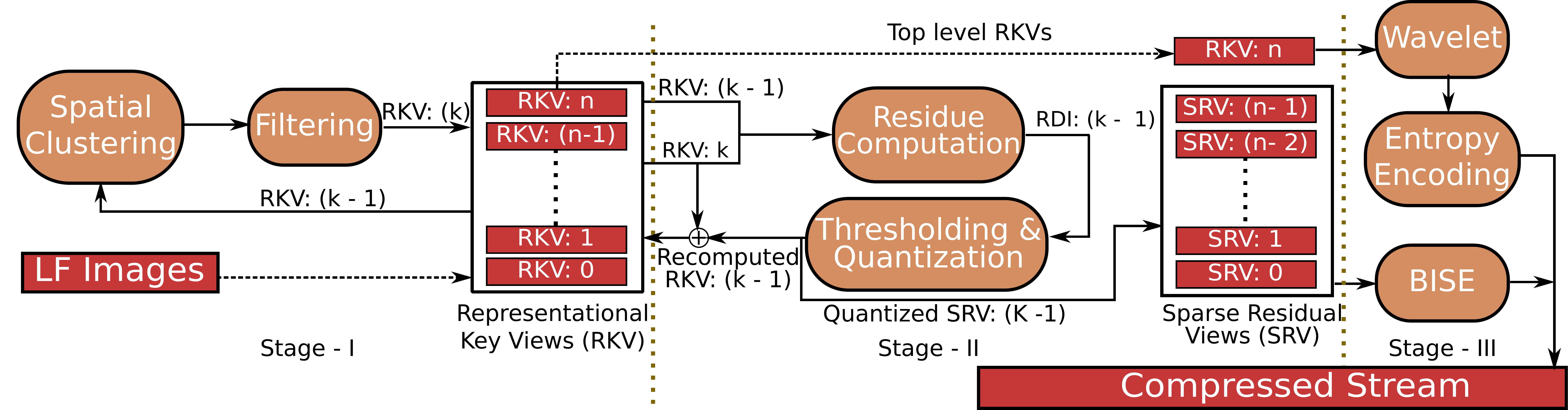}

 \caption{Our compression pipeline: The red rectangles represent data, the brown ovals represent processing blocks, and the arrows indicate flow and data transfer operations. The compression pipeline is comprised of three stages. (Stage - I) The first stage computes hierarchical representational key views (RKVs), where each level is computed by filtering clusters from the previous level. (Stage- II) The second stage consists of computing sparse residual views (SRVs) through a top-down approach, starting at the top level. (Stage - III) In the third stage, the top level RKV  and SRV levels are further processed to compute the compressed bitstream. }
 \label{fig:comp_pipeline}
 \vspace{-0.75em}
 \end{figure*}

\subsection{Light Field Rendering}

The intensity of light rays through empty space remains constant. ~\citet{LFLevoy96} use this observation to present a simple two-plane (4D) parametrization of the plenoptic function
and describe a practical light field rendering scheme for real-time photo-realistic rendering of complex objects. In this two-plane parameterization, all the light rays between parallel planes are described using a pair of parameters $(u, v)$ and $(s, t)$. The light rays between the two planes are called \textit{light slab} and the 4D parametrized plenoptic function is called \textit{light field} (LF). Gortler et al. [1996] presents a different representation of the 4D parameterization called a ~\textit{lumigraph}. In a lumigraph, a cube is used instead of two parallel planes to  create a bound over a particular region of interest in
space. The light rays in any light field parameterization~\citep{LFSphereihm1997,LFdavis2012unstructured} are captured using a very large set of discretely sampled camera images. This large amount of data required creates a practical bottleneck for capturing LFs  and for rendering methods. 

\subsection{Light Field Compression}
The minimum sampling rate required for a good reconstruction using IBR is a well-studied problem~\cite{LFsampchai2000sampling,LFsampchan2000sampling}. Even   with a minimal sampling rate, the number of image samples required can be in the order of thousands for a good-quality reconstruction~\cite{LFsampchai2000sampling}. Many schemes have been proposed for LFI compression and we categorize the existing compression schemes into two categories; \textit{high efficiency encoding schemes}, which include methods similar to standard image and video coding techniques (DCT, wavelet); and \textit{random access compression schemes}, which include methods suitable for fast viewing and rendering as they provide fast random access to specific pixels, but provide lower compression ratios. In response to the growing interest in quality plenoptic content , the JPEG standardization committee launched JPEG Pleno~\cite{JPEGPleno}. The aim of  JPEG Pleno is to define standards for the wide adaptability of 4D LF compression, like JPEG and MPEG standards.
 
\subsubsection{High efficiency LFI compression schemes:}

Earlier work on LFI compression is based on extending  standard image and video coding methods (JPEG, JPEG200, MPEG-2, MPEG-4). These methods employ techniques such as discrete-cosine transform, wavelet transform, predictive block encoding, and motion-vector compensation.~\citet{LFmagnor2000,LFGirod03, LFjagmohan2003compression} use disparity compensation instead of motion vectors for predictive coding of blocks of LFI. In disparity compensated approaches, a pre-fixed set of LF images is encoded independently (I-frames) and the rest of the LF image blocks are encoded predictively (P-frames) from the I-frame blocks. Due to the uniform camera motion in the sampling of LF in two plane parameterization, P-frame blocks can be predicted from I-frame blocks using a single disparity value. The compression rates of these methods are around 100:1 to 200:1, depending on the details in the original LFI.  ~\citet{LFKundu12} uses homography techniques to predictively encode LF images (P-frames) by warping them onto a set of I-frames, achieving compression rates of 10:1 to 50:1. Chang et al.~\citeyear{LFchang2006light} use techniques that are based on using additional shape and geometry information about the object captured in the LF images. More recently, methods directly based on HEVC video coding have been proposed by~\citet{LFliu16pseudo,LFperra2016high,LFchen18} and they can achieve high compression rates of 100:1 to 1000:1. ~\citet{LFliu16pseudo}  order the LF-Images using a pseudo-sequence temporal ordering and compress them using HEVC encoding. ~\citet{LFchen18} use a small set of views to predict the rest of the images using disparity based image-transformations and combine it with a pseudo-sequence method~\cite{LFliu16pseudo}. 

\subsubsection{Random Access LFI compression:} 
 ~\citet{LFLevoy96} present a compression technique using vector-quantization (VQ) that provides random access for interactive rendering. VQ results in compression rates of around 10:1 to 20:1, but the compression quality is low. Moreover, VQ-based compression fails to take advantage of the high correlation or coherence between LF images. ~\citet{LFPeter01} describe an approach for random-access compression using a 4D wavelet hierarchical scheme, which provides compression rates of 20:1 to 40:1. However, this method requires multi-level caches for fast data access during rendering. This approach also makes assumptions about the scene captured in the light fields. ~\citet{zhang2000compression} describe a technique that is similar to the high-efficiency compression schemes using multi-reference frame-based motion compensation  and achieves compression ratios of 80:1. ~\citet{WelcomeLF} present a new end-to-end system for capturing and rendering very high resolution light fields with a new capturing system, a new spherical parameterization, and a new rendering approach for light fields. ~\citet{WelcomeLF} present a compression scheme using motion compensated prediction by modifying the VP9 video codec~\citep{VP9} to allow for more reference frames and provide random access. They achieve compression rates of 40:1 to 200:1 for a high quality compression. ~\citet{KoniarisDisney} describe an end-to-end system for rendering animated light fields. They present a temporal compression scheme that uses interpolation between the frames (in time dimension) to encode the animated light fields. On top of temporal compression, they utilize standard texture compression methods (DXT, BC6, BC7) to achieve more gains in compression. Overall they achieve  compression ratios of 60:1 to 500:1 (temporal and spatial) with a spatial compression ratio of 6:1.
 

\vspace*{-1em}

\section{Our Method: RLFC}
\label{sec:method}


In this section, we present an overview of our compression pipeline (Figure~\ref{fig:comp_pipeline}) and the details of our encoding method. The input to our method is sampled light field images of a scene and the output is a compressed stream that provides parallel decoding capabilities and random access. 
Our approach has the following components:


\vspace*{0.1in}

\noindent  \textbf{Representative Key Views} (\textbf{RKVs}): At each level of the tree, these 2D images capture the redundancies present in the images in the level below it. In the final stream, only the top-level \textbf{RKVs} are stored.

\vspace*{0.1in}

\noindent \textbf{Sparse Residual Views}: (\textbf{SRVs}): 
These 2D images are stored at each level and capture the specific details of the images in the current level.

\vspace*{0.1in}

\noindent \textbf{Clusters} (C): Clustering is performed to gather the LF samples (2D-images sampled while capturing the LF) that are spatially close to one another (localized cluster computation) and which exhibit coherency. We cluster the RKVs that are close one another at each level of the tree. Samples refer to the 2D images sampled while capturing the light field.

 \vspace*{0.1in}

\noindent \textbf{Filtering} : Filtering is performed to compute a single representative view that captures the redundancies of the samples in a cluster. The representative key view at the next level is computed for samples that are close to one another using a weighted filtering of the samples in each cluster. 

\vspace*{0.1in}


\noindent \textbf{Residue Computation, Thresholding, \& Quantization}: The residue computation is performed to compute the \textbf{SRV} in a top-down fashion by taking the difference between alternate levels of \textbf{RKVs}, followed by thresholding and quantization of the \textbf{SRVs} at that level. The thresholding step is performed to remove the insignificant residual blocks in \textbf{SRVs}. The insignificant blocks are determined by a threshold set as an encoding parameter. The block size is also set as an encoding parameter. After thresholding the pixel values in the significant blocks are quantized to reduce the dynamic range of the pixel values in \textbf{SRVs}.

\vspace*{0.1in}

\noindent \textbf{JPEG200}: All the top-level $\mathbf{RKVs}$ are compressed in a lossless mode and we use the JPEG2000 algorithm.

\vspace*{0.1in}

\noindent \textbf{BISE}: \textit{Bounded Integer Sequence Encoding} (BISE) is a method used to encode a sequence of integer values in different ranges in an efficient hardware-friendly manner. It is used for texture compression and we use BISE to encode  the integer values in the quantized \textbf{SRV} blocks. It provides good compression rates and is well supported on current GPUs.

\begin{figure}[t!]
 \centering
 \includegraphics[width=\columnwidth, keepaspectratio=true]{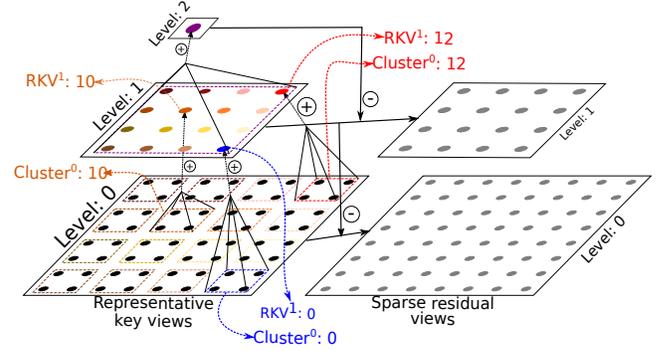}

 \caption{We highlight the construction of hierarchical representative key views (RKVs) and sparse residual views (SRVs) for two plane LF parametrization. Level:0 corresponds to the original LF image samples. Level: 1 is computed by filtering clusters of four spatially close images. Level: 2 is constructed by Gaussian weighted filtering of the images in the clusters on level: 1. Sparse residual views (SRVs) are constructed in a top down manner by computing the differences between alternative levels of representational views. The subscript in the figure indicates the level and the number following indicates the indices of both the cluster and the RKV on the current level.}
 \label{fig:level}
 \vspace*{-1.5em}
 \end{figure}

\vspace*{0.1in}
We construct a tree with the root node storing representative key views (\textbf{RKVs}) and the children nodes storing the sparse residual views (\textbf{SRVs}). Our construction process starts by setting the input indexed LFI as the bottom level (zero level) of the tree and then we recursively construct higher-level RKVs starting from the bottom level to the top level. At any given level, we cluster the RKV images that are spatially close; for all the clusters, we compute the RKVs of the next level of the tree. After the RKV tree is computed,  we compute the SRVs for each level  from top to bottom, reconstructing RKV images after the thresholding and quantization of SRVs at each level of the tree. After the SRVs are computed, only the top level RKVs and the SRVs computed are stored (Fig. ~\ref{fig:level}). Next, we encode the RKV images at the top level (root) of the tree using standard image compression techniques such as JPEG200. Each level of the SRV is  divided into blocks and compressed using Bounded Integer Sequence Encoding (BISE)~\cite{ASTC}. 

\textbf{Notation:} We use the following short forms and notation while presenting our approach: 
$\mathbf{RKV}^{l}$ denotes the set of all representative key views at level $l$ of the tree; $(RKV_{i})^{l}$ denotes the $i^{th}$ representative key view at level $l$; 
$\mathbf{SRV}^{l}$ denotes the set of all sparse residual views at level $l$ of the tree; $(SRV_{i})^{l}$ denotes the $i^{th}$ sparse residual view at level $l$; $(C_{j})^{l}$ denotes the $j^{th}$ set of clustered $RKV$ samples on the level $l$. 

\subsection{Representative Key Views} 
The image samples in an LF exhibit strong spatial correlation and we exploit this to find redundancies and compress the data. The first step is to cluster samples that are close to each other using the number of clusters and the number of images in each cluster set as an encoding parameters for all the levels:
\newline
\begin{eqnarray}
 C_j^{(l - 1)} = \bigcup RKV_k^{(l - 1)} \\ \nonumber
k \in \lbrace k_1, k_2, ... k_n  \mid dist(k_u, k_v) < threshold \rbrace,
\end{eqnarray}
where $k$ denotes the index $RKV$ image on level $(l - 1)$. $dist(k_u, k_v)$ computes the spatial distance between $RKV_{k_u}$ and $RKV_{k_v}$ at current level. Spatial distance corresponds to the Euclidean distance between the spatial locations (relative or absolute real-world) of the LFI provided in the input. For each cluster of image samples, an $RKV$ image is computed. The $RKV$ image is computed using a weighted filtering of all the images in a given cluster. Let $(C_j)^{(l - 1)}$ be the $j^{th}$ cluster on level $(l - 1)$, $I$ denote an $RKV$ in the current cluster and the $(RKV_j)^{l}$ on level $l$ is computed as:

\begin{equation}
RKV_j^{l} = \sum_{I\in C_j^{(l - 1)}}w_{jI}^{(l - 1)}\ \times\ I.
\end{equation}

\begin{figure}[t!]
 \centering
 \includegraphics[width=\columnwidth, keepaspectratio=true]{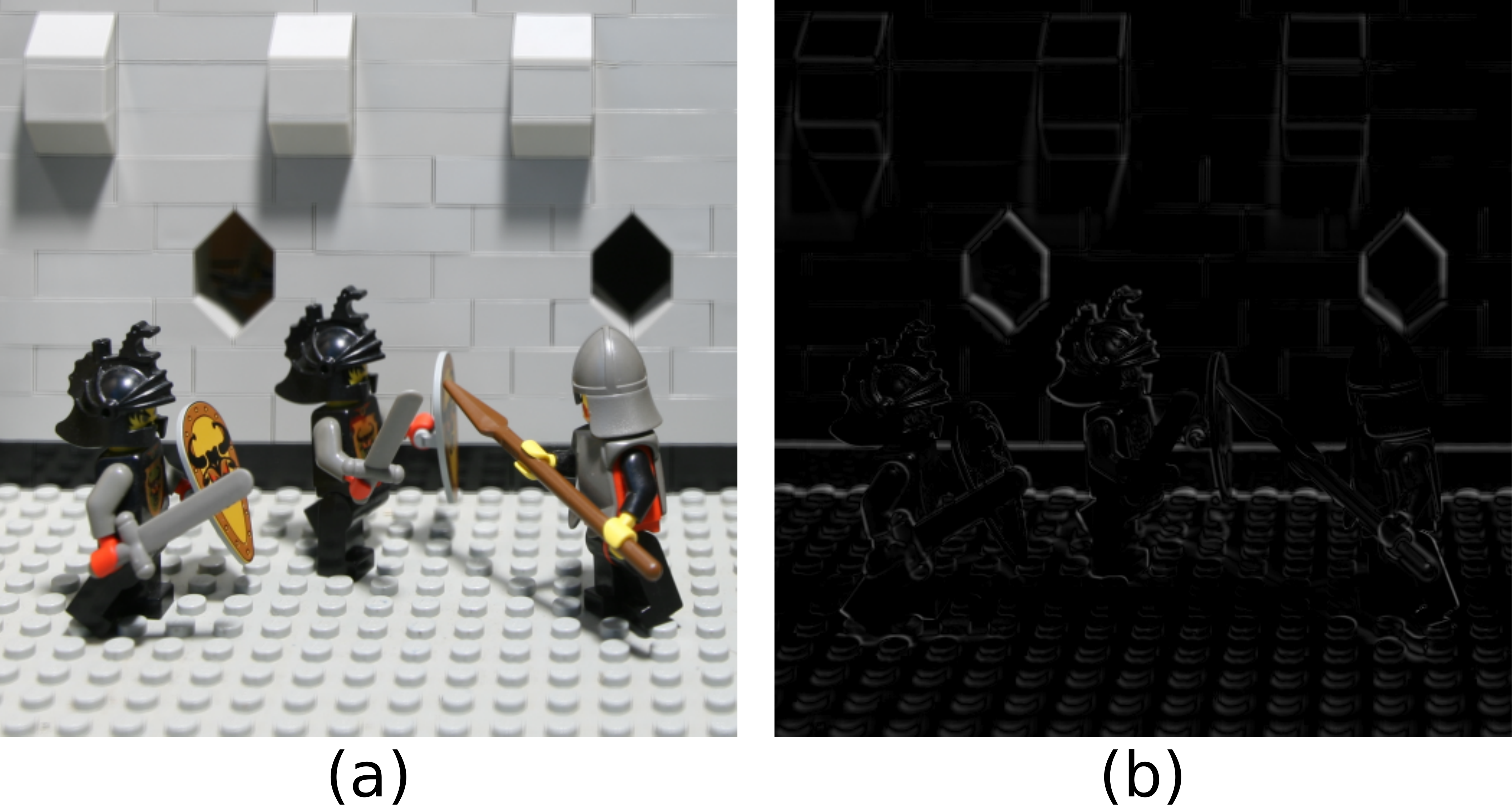}
 \caption{(a) Example of a representative key view (RKV) and (b) a sparse residual view image (SRV). We can see that the residual view image is very sparse with large intensity pixel values constrained in small regions. The images are generated from the Lego Knights LF from the Stanford light field archive.}
 \label{fig:rv_rdi}
\vspace*{-1em}
 \end{figure}

$I$ denotes the images in the cluster $C_j^{(l - 1)}$. $w_{jI}^{(l -1)}$ is the weight for image $I$ for filtering images in the cluster $C_j^{(l - 1)}$  The new set of $\mathbf{RKV}^{l}$ images computed at a level $(l)$ exhibit spatial correlations similar to the $\mathbf{RKV}^{(l -1)}$ images at the level below $(l - 1)$. This process is recursively repeated until a certain number of levels of the tree is computed. The number of levels is one of the   encoding parameters. This process generates hierarchical levels  of $RKVs$, which results in a child-parent relationship between the cluster of images $C_j^{(l -1)}$ at level $(l - 1)$ and the corresponding $RKV_j^{l}$ image at level $l$. Figure~\ref{fig:level} illustrates this process for a given two-plane parameterization. We use weighted Gaussian filtering in our analysis to capture the redundancies among the spatially close by RKV. Another advantage of using filtering is that it is very fast to compute and enables real-time encoding inexpensively. 


\subsection{Sparse Residual Views}

Once the $RKV$ tree is constructed, the $SRV$ tree is computed in a top-down manner, starting with the top-level of the $RKVs$ tree.  The $SRV^{(l - 1)}$ at level $(l - 1)$ is computed by subtracting the parent $(RKV_p)^{l}$ at level $l$ from the corresponding children $RKV^{l - 1}$  at level $l -1$:
\begin{equation}
(SRV_i)^{(l - 1)} = (RKV_{i})^{(l - 1)} - (RKV_p)^{l}.
\end{equation}
The $SRV$ images tend to be quite sparse and consist of regions of high intensity values. Figure~\ref{fig:rv_rdi} shows an example $RKV$ and $SRV$ computed using our approach. It can be seen that the $SRV$ image~\ref{fig:rv_rdi} is mostly empty with zero or small intensity and only certain regions have high intensity pixel values. Based on this observation, we divide all the levels of the $SRV$ images into non-overlapping rectangular blocks and only store the blocks with high intensity pixel values. The $SRV$ images are processed using a thresholding step followed by a quantization step. We process the $SRV$ images with two levels of thresholding, pixel level thresholding and block energy thresholding, which are based on thresholds set as encoding parameters. The pixel level thresholding determines whether pixel values in a $SRV$ image provides a sufficient contribution. If pixel values are below the set threshold the pixel is considered insignificant and the value is set to zero. The block energy based thresholding computes the sum of the absolute pixel values in a block and decides whether a block has block energy above the set threshold. If the block energy is below the set threshold, the block is considered insignificant and discarded. After thresholding, the dynamic range of the pixel values in the significant blocks is reduced by quantizing the pixel values. More details on the quantization are presented in the suppl. material, Section-3 \footnote{supplementary material can be found at: \url{https://bit.ly/2MwEGjv}}. Once the $\mathbf{SRV}^{(l - 1)}$ at level $(l - 1)$ is computed, the $\mathbf{RKV}^{(l - 1)}$ at the corresponding level is recomputed. We recompute $\mathbf{RKV}^{(l - 1)}$ so that the thresholding and quantization errors do not propagate down the tree to the levels below. Once the downward pass is finished, the $\mathbf{RKVs}$ at all levels are discarded except for the top level (n) $\mathbf{RKV}^n$. In the end we are left with a tree with the root as the top-level $\mathbf{RKV}^n$ and the children nodes are the sparse quantized $\mathbf{SRV}$ levels. The $\mathbf{RKV}^n$ at the root of the tree is compressed using standard image compression techniques (JPEG2000). An example visualization of the computed trees (SRV images and RKV images) is presented in  suppl. material, Section-5.

 \begin{figure}[t!]
 \centering
 \includegraphics[width=\columnwidth, keepaspectratio=true]{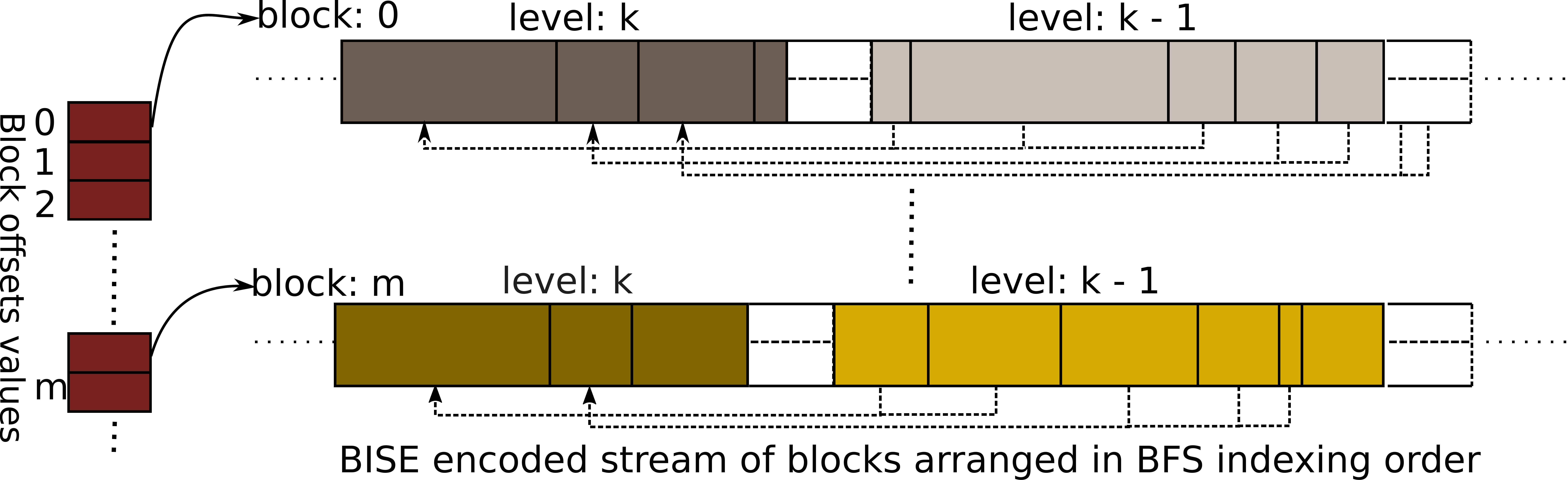}

 \caption{The arrangement of the SRV tree in the final compressed stream. All the blocks at the same particular spatial location are gathered from all the nodes of the SRV tree. The  BISE encoded streams of the gathered blocks are appended to the compressed stream in the serial order computed using BFS traversal. For each block, the start locations of the compressed stream are stored in a block offset array, shown on the right.}
 \label{fig:comp_stream}
\vspace*{-2em}
 \end{figure}

\subsection{Bounded Integer Sequence Encoding}
BISE was first introduced and used in the ASTC texture compression format by~\citet{ASTC}. For a set of integer values that lie in the range $0$ to $N - 1$ with equal probability, BISE addresses the problem of encoding them efficiently while allowing constant time decoding with a very limited hardware. Apart from trivial case of storing $\log_{2}{N}$ bits, when $N$ is a power of two, BISE describes an efficient packing method for different ranges of $N$ as well. We use BISE to  encode the SRV blocks for two reasons:
\begin{enumerate*}
\item Traditional techniques like DCT or wavelet transform separate a given signal into high-frequency coefficients and low-frequency coefficients. We observe that $\mathbf{SRVs}$ are high-frequency signal images and that using traditional techniques does not work well (suppl. material, Section-1).
\item BISE provides good compression rates with hardware supported decoding available on most desktop and mobile commodity GPUs. 
\end{enumerate*}

 \begin{algorithm}
\caption{Compress light field image samples}\label{alg:comp}
\begin{algorithmic}
\Require 
\Statex Original indexed light field images: \textbf{LFI}
\Statex Encoding parameters: \textit{enc}
\Ensure 
\Statex Compressed stream 
\Function{CompressLFI}{$\mathbf{LFI}, \textit{enc}$}
\State //\textcolor{gray}{Set the bottom level of RKV tree to LFI}
\State $\mathbf{RKV}^{0}\gets \mathbf{LFI}$
\State $\mathbf{RKV} \gets ComputeRKVTree(\mathbf{RKV}^0, enc)$\
\State $\mathbf{SRV} \gets ComputeSRVTree(\mathbf{RKV}, enc)$
\State //\textcolor{gray}{Compress top level $RKV^n$ using JPEG2000}
\State RKVStream $ \gets JPEG2000(\mathbf{RKV}^{n})$
\State \textcolor{gray}{//Initialize the SRV tree compressed bitstream to empty}
\State SRVStream $\gets 0$
\State BlockOffsets $\gets \lbrace\rbrace$
\State //\textcolor{gray}{Process all the blocks}
\ForEach {index $in$ blocks}
\State \textcolor{gray}{/*Travel the SRV tree in a level order fashion 
\State to gather the blocks*/}
\State OrderedBlocks $\gets BFS(\mathbf{SRV}, index)$
\State \textcolor{gray}{//Encode the blocks of SRV levels using BISE}
\State OrderedBlocksBISE $\gets BISEEncode($LvlBlocks$)$
\State BlockOffsets $\gets$ Sizeof(OrderedBlocksBISE)
\State SRVStream $\gets$ SRVStream$\colon$OrderedBlockBISE
\EndFor
\State \textcolor{gray}{//Append the streams and return the final stream}
\State  \textbf{return} (RKVStream$\colon$BlockOffsets$\colon$SRVStream)
\EndFunction
\end{algorithmic}
\end{algorithm}

\subsection{Compressed Stream Structure}
We arrange the final compressed stream to enable progressive and random access decoding of the pixels. To create a compressed stream, the final tree (top-level $\mathbf{RKV}^n$ and $\mathbf{SRV}$) must be linearized. The first step in creating our compressed stream structure is assigning a serial indexing (linearization) to all the nodes in the final tree. We traverse the final tree from the root using breadth first search (BFS), indexing all the nodes in the order they are traversed by BFS.

After the $\mathbf{SRV}$ nodes are linearly ordered using BFS, all the blocks of each SRV node are assigned the same serial index as the node. BISE encoded blocks in the same spatial location are gathered from all the nodes in the $\mathbf{SRV}$ tree and appended to the compressed stream in the corresponding serial order from BFS. In this manner, all the BISE encoded blocks of the $\mathbf{SRV}$ are processed starting with the block on the top-left and ending with the block on the bottom-right. An array of block offset values to the start location of each block's compressed stream in the final stream is stored to facilitate parallel and random-access decoding. Figure~\ref{fig:comp_stream} highlights our compressed representation.

 \begin{algorithm}
\caption{Decompress light field image block}\label{alg:decomp}
\begin{algorithmic}
\Require 
\Statex LFI compressed stream: \textbf{CompLFI}
\Statex Image index: ImgIdx
\Statex Block index: {BlkIdx}
\Ensure 
\Statex Pixel values: PixVals
\Statex \textcolor{gray}{// Load the stream into memory and separate }
\Statex \textbf{Initialization:}
\Statex  RKVStream $\gets$ $ReadRKVStream(\textbf{CompLFI})$
\Statex  $\mathbf{RKV}^{n} \gets$ $DecompressJPEG2000($RKVStream$)$
\Statex BlockOffsets $\gets$ \textit{ReadBlockOffsets}$(\textbf{CompLFI})$
\Statex SRVStream $\gets$ $ReadSRVStream(\textbf{CompLFI})$
\Statex
\Function{DecompressLFIBlock}{ImgIdx, BlkIdx}
\State \textcolor{gray}{// Get the start location of BlkIdx in bitstream}
\State StartOffset $\gets$ BlockOffsets[BlkIdx]
\State \textcolor{gray}{// Read the top level filtered values from \textbf{RKV}}
\State RKVBlock $\gets ReadBlock(\mathbf{RKV}^{n}$, BlkIdx $)$
\State \textcolor{gray}{// Compute the location of the parent blocks in stream}
\State ParentIndx $\gets$ $GetParentIndices($ ImgIdx $)$
\State \textcolor{gray}{//Read the BISE encoded stream of required blocks }
\State OrderedBlockBISE $\gets ReadBlocks($SRVStream, ParentIndx$)$
\State \textcolor{gray}{//Decode the BISE blocks}
\State OrderedBlocks $\gets BISEDecode($OrderedBlockBISE$)$
\State \textcolor{gray}{//Combine the residual values with filtered pixel values}
\State PixVals $\gets CombineBlocks($RKVBlock, OrderedBlocks$)$
\EndFunction
\end{algorithmic}
\end{algorithm}

\subsection{Decompression \& Random Access}

The first step in decompression is decoding all the top-level $RKV$ images. To decode a block of pixels at a particular location, we traverse the $SRV$ tree and gather the BISE streams of all the corresponding blocks until the process reaches the bottom level. Next, the residual pixel values are gathered by decoding the BISE streams. Finally, the pixel values are computed by combining the residual pixel values with the corresponding top-level $RKV$ image. While rendering, only a small set of pixels from a different LFI is required by the renderer. Using the tree traversal operation, we selectively gather and decode only the blocks required by the LF renderer. With selective decoding, our compressed stream stays the same in the memory. Only small parts of the compressed stream are decoded to get the pixel values required by the LF renderer.

Given a pixel's location, we can compute the individual block location and the location of all parent-child blocks in the compressed stream by running the same BFS used during compression as an initialization step of decompression. The start location of the current block's compressed stream is located using the block offset values. Using the tree traversal decompression, our approach provides random access to the pixel values at block level without decompressing the other parts of the compressed data.

\subsection{Interactive Rendering}
Our decompression scheme is designed to be compatible with several efficient IBR techniques~\cite{IBRisaksen2000dynamically,LFLevoy96} because it supports block-based parallel and progressive decompression. It is evident from the arrangement of the compressed stream (Figure~\ref{fig:comp_stream}) that our method supports parallel decoding of multiple blocks at once. Our method inherently supports progressive decompression because the $RKV$ tree constructed during the compression has filtered images of spatially close by LF images at different levels. For progressive decompression we can modify the tree traversal operation to stop and compute the high-level filtered pixel value at a certain level. The rendering scheme can progressively update the pixel values, as the decompressor computes the final pixel values in the background. Parallel and progressive decompression can significantly improve the rendering rate at runtime. The decompression scheme is also hardware friendly because the decoding operations consist of memory reading and simple integer arithmetic operations. 

During rendering, the pixel values of the new views are computed by interpolating several surrounding pixel values from several LF samples. Our method inherently supports the efficient reconstruction of new views by interpolating multiple samples. When we decode a pixel value, an entire block is decoded, which also provides surrounding pixel values. Therefore, while generating new views, a set of spatially nearby blocks from the LF samples is decoded. The intensity of the new pixel value ((u, v, s, t) ray in 4D parameterization) is computed by quadrilinear or bilinear interpolation of several pixel values from the decoded blocks.

During interactive rendering, the renderer requests pixels from several locations of the light field images. The GPU decompression algorithm decodes blocks of the LFI in parallel to get the required pixel values. The first step of decoding involves reading the required compressed data from memory. The compressed data in the memory is arranged in a BFS linearization representation of the computed hierarchy (Fig.~\ref{fig:comp_stream}). The memory access pattern during the first step of memory reading corresponds  to \textit{gather-pattern}, as the decoder needs to read from different random blocks of compressed data from the linearized tree representation in the memory. The memory-access pattern while writing the decoded data corresponds to a \textit{sequential write-pattern} as the decompressed pixel values are written into a fixed index memory buffer allocated for the new view.

{\small
\begin{table}[t!]
\centering
\begin{tabular}{|c|c|c|}
\hline
LF Dataset (Resolution): Size (MB) & \begin{tabular}[c]{@{}c@{}}Compression  \\ rate (bpp)\end{tabular} & PSNR (dB) \\ \hline
Dragon $(32 \times 32 \times 256 \times 256): 192$                                                  & 0.290                                                              & 41.39     \\ \hline
Budhha $(32 \times 32 \times 256 \times 256): 192$                                                 & 0.084                                                              & 41.21     \\ \hline
Amethyst $(16 \times 16 \times 768 \times 1024): 576$                                               & 0.109                                                              & 41.99     \\ \hline
Bracelet $(16 \times 16 \times 1024 \times 640): 480$                                               & 0.542                                                              & 41.31     \\ \hline
Bunny  $(16 \times 16 \times 1024 \times 1024): 768$                                                & 0.094                                                              & 40.85     \\ \hline
Jelly Beans $(16 \times 16 \times 1024 \times 512): 384$                                            & 0.172                                                              & 40.95     \\ \hline
Lego Knights $(16 \times 16 \times 1024 \times 1024): 768$                                          & 0.62                                                               & 40.64     \\ \hline
Lego Gallantry $(16 \times 16 \times 640 \times 1024): 480$                                         & 0.40                                                               & 40.15     \\ \hline
Tarot Cards $(16 \times 16 \times 1024 \times 1024): 768$                                           & 2.20                                                               & 41.99     \\ \hline
\end{tabular}
\caption{The compression rates and quality for several LF datasets from the Stanford light field archive. All the image samples are 24-bit color RGB images. For a similar PSNR quality, the compression rate varies for each LF depending on the details of the scene recorded in the LF.}
\label{tab:overall}
\vspace*{-2em}
\end{table}
}

\subsection{Performance Analysis}
Algorithm-~\ref{alg:comp} gives a high level pseudo-code of our compression scheme. The primary operations involved in  our compressing scheme are: \begin{enumerate*}
\item Filtering images;
\item Compressing the top-level $\mathbf{RKVs}$ of the computed tree;
\item Rearranging blocks of memory, $\textbf{SRV}$ Breadth first search order traversal;
\item BISE encoding of sparse residual blocks.
\end{enumerate*} 

Algorithm-~\ref{alg:decomp} highlights the steps in the decompression scheme. At the start of the rendering operation, the final compressed stream is loaded into the memory and the top-level $RKV$ images are decompressed. The operations used in decoding a block computation include: \begin{enumerate*}
\item loading of required bytes from in memory SRVStream into registers;
\item bit manipulation operations required for decoding BISE compressed blocks;
\item simple integer arithmetic operations to combine the SRV pixel values with the RKV pixel values; 
\end{enumerate*}. Our decompression scheme is hardware friendly because the operations include memory reads, bit manipulation, and simple integer arithmetic.

\section{Evaluation \& Analysis}
\label{sec:results}
We have implemented RLFC for the two plane parameterized LF. We have tested and analyzed our approach on the Stanford light field archives~\cite{LFDatawilburn2005high,LFLevoy96}. The LFIs in the dataset from~\citet{LFDatawilburn2005high} are high resolution images captured using large camera arrays. The input LF images are $24$-bit RGB images. We use lossless YCoCg-R~\cite{InvertibleYCoCg} color space to decorrelate the color channels. During all stages of our compression scheme, we use lossless integer computations. We measure the compression rate using bits per pixel (bpp). In the current implementation, we compress the top level representative views using the JPEG2000 lossless algorithm. The quality is measured using \textit{peak-signal-to-noise-ratio}(PSNR\textsubscript{YCoCg}) as the weighted average~\cite{PSNROHM} of the PSNR individual components:
\begin{equation}
PSNR_{YCoCg} = \frac{6\ *\ PSNR_{Y} + PSNR_{Co} + PSNR_{Cg} }{8}
\end{equation}
The PSNR of each component is measured dB using:
\begin{equation}
PSNR = 10\ * \log_{10} \frac{255^2}{MSE}
\end{equation}

MSE is the mean square error between the original images and the decompressed images in the LF. The final PSNR is computed as the average of all the images in the LF.

\begin{table}[t!]
\begin{tabular}{|c|c|c|c|c|}
\hline
LF Dataset & Metric & \begin{tabular}[c]{@{}c@{}}Tree\\ Height: 3\end{tabular} & \begin{tabular}[c]{@{}c@{}}Tree\\ Height: 4\end{tabular} & \begin{tabular}[c]{@{}c@{}}Tree\\ Height: 5\end{tabular} \\ \hline 
\multirow{2}{*}{Amethyst} & bpp & 0.265 & 0.224  & 0.221\\ \cline{2-5}
                           & PSNR & 44.69 & 44.28 & 44.03\\ \hline \hline
\multirow{2}{*}{Bunny} & bpp & 0.227 & 0.180  & 0.171\\ \cline{2-5}
                      & PSNR & 44.69 & 44.33 & 44.00\\ \hline \hline
\multirow{2}{*}{Bracelet } & bpp & 0.822 & 0.809 & 0.81 \\ \cline{2-5}
                           & PSNR & 45.89 & 45.11 & 44.81 \\ \hline \hline 
\multirow{2}{*}{Knight} & bpp & 0.731  & 0.670 & 0.657 \\ \cline{2-5}
                           & PSNR & 43.85 & 43.48 & 43.31 \\ \hline    
\end{tabular}
\caption{The variation in the resulting compression rate and quality varies with changes in \textit{tree height} is highlighted. The \textit{block threshold} is set to 80, the \textit{block size} is set to 4, and \textit{quantization level} is set to 2 for all the datasets under consideration. As the \textit{tree height} is increased, the sparsity of the residual levels in the tree increases. With the \textit{block threshold} fixed the thresholding errors increase and there is a  slight decrease in the resulting PSNR and bpp.}
\label{tab:treeht}
 \vspace*{-2em}
\end{table}



Table~\ref{tab:overall} shows the compression rate and PSNR for several LF datasets for a \textit{tree height} of 3. The encoding parameters are adjusted for each dataset to achieve a similar decompression quality. The compression rate varies from $0.08 - 2.5$ bpp for a similar PSNR quality, depending on the details of the scene captured in the LF. 


We measure the effect of encoding parameters on the resulting compression rate, and quality. Table~\ref{tab:blksize} highlights the effect of changing the \textit{block size} on the bpp and PSNR with the \textit{block threshold} and the \textit{tree height} set as constants. For a fixed \textit{block threshold} and \textit{quantization level}, increasing the \textit{block size} increases the energy (the absolute sum of pixel values) of the residual blocks, reducing the thresholding errors and resulting in higher bpp and PSNR. In Table~\ref{tab:treeht}, we study the variations in the resulting bpp and PSNR with a change in the \textit{tree height}. As the \textit{tree height} increases, the sparsity of the residual views in each level increases. With a fixed \textit{block threshold}, the thresholding errors are increased with an increase in sparsity, and  slight reduction in the bpp and quality are observed. 

\begin{figure*}[t!]
 \centering
 \includegraphics[width=\textwidth, keepaspectratio=true]{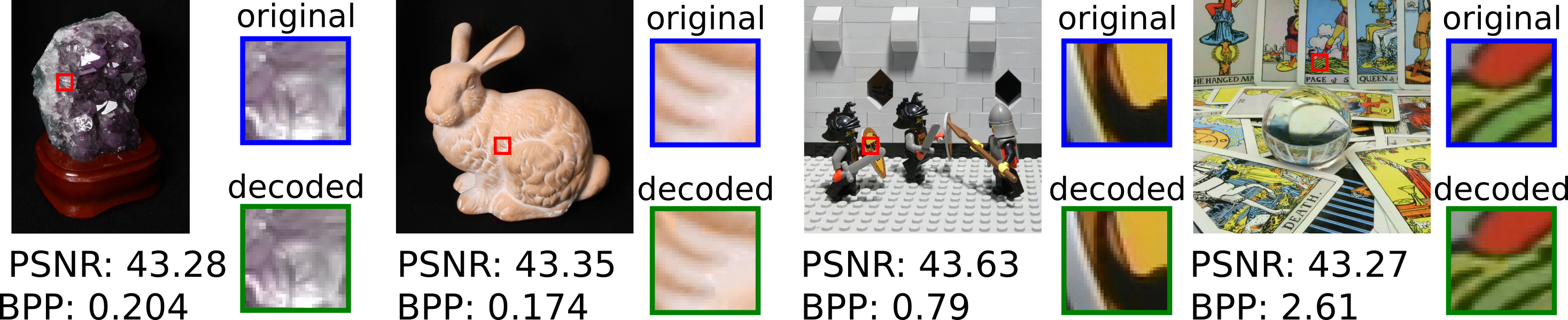}

 \caption{ The zoomed in comparisons between original and decoded LF images from our method are shown. $32\times32$ small regions as highlighted in red boxes are scaled to $512\times512$ to compare the visual quality between original and compressed images. We observe that the visual quality is not degraded in the images compressed using our method.}

 \label{fig:zoomed_in}
 \vspace*{-1em}
 \end{figure*}

\begin{figure}[t!]
 \centering
 \includegraphics[width=\columnwidth, keepaspectratio=true]{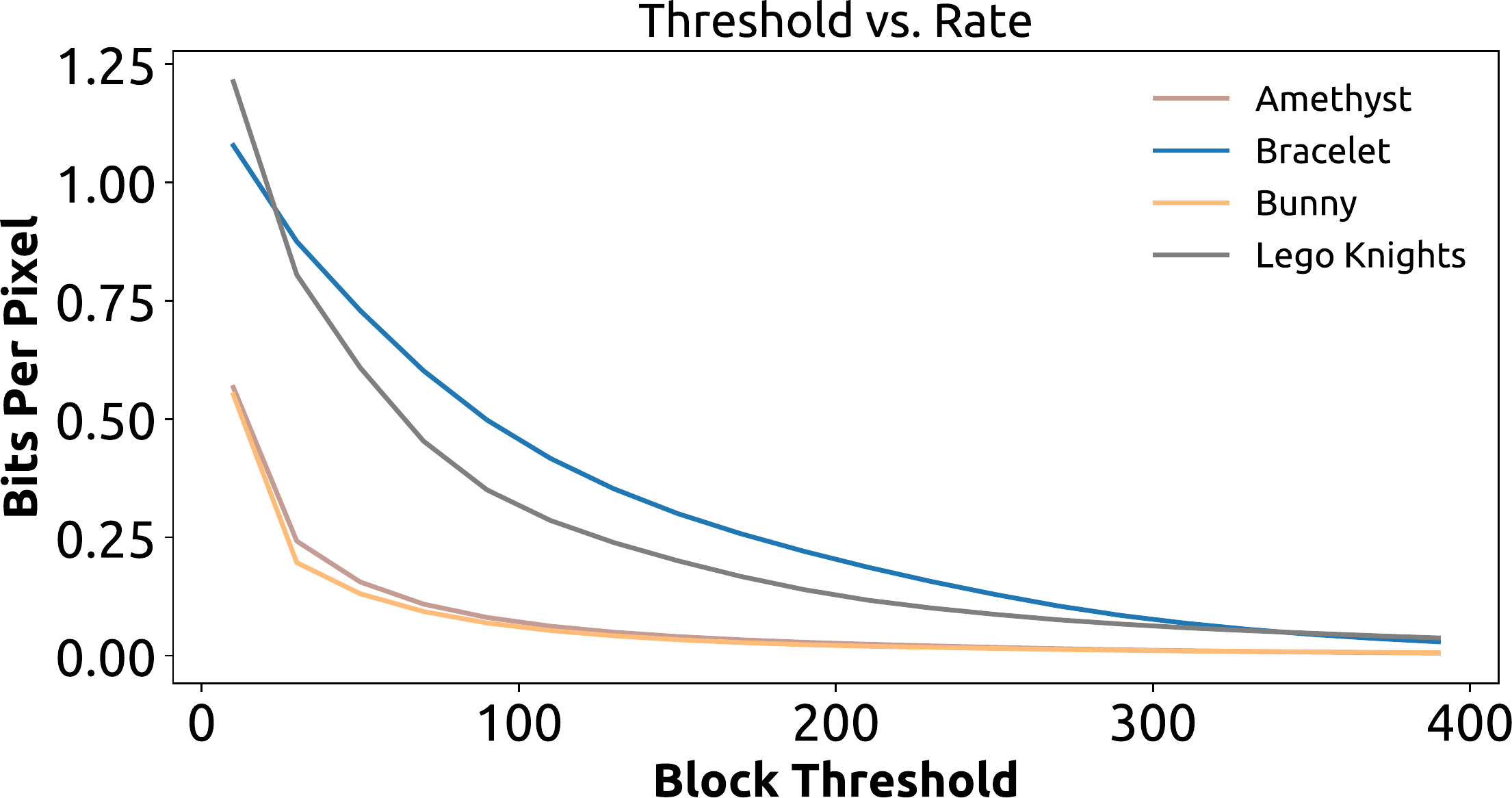}

 \caption{The variation of the compression rate (bpp) with change in the block threshold is plotted. The encoding parameters are set as: \textit{tree height} to 3, \textit{block size} to 2, \textit{quantization level} to no quantization, and the \textit{block threshold} is varied. With an increase in the \textit{block threshold} the thresholding errors and the bpp decreases for all the LF datasets.}
 \label{fig:thrsh_bpp}

 \end{figure}

\begin{figure}[t!]
 \centering
 \includegraphics[width=\columnwidth, keepaspectratio=true]{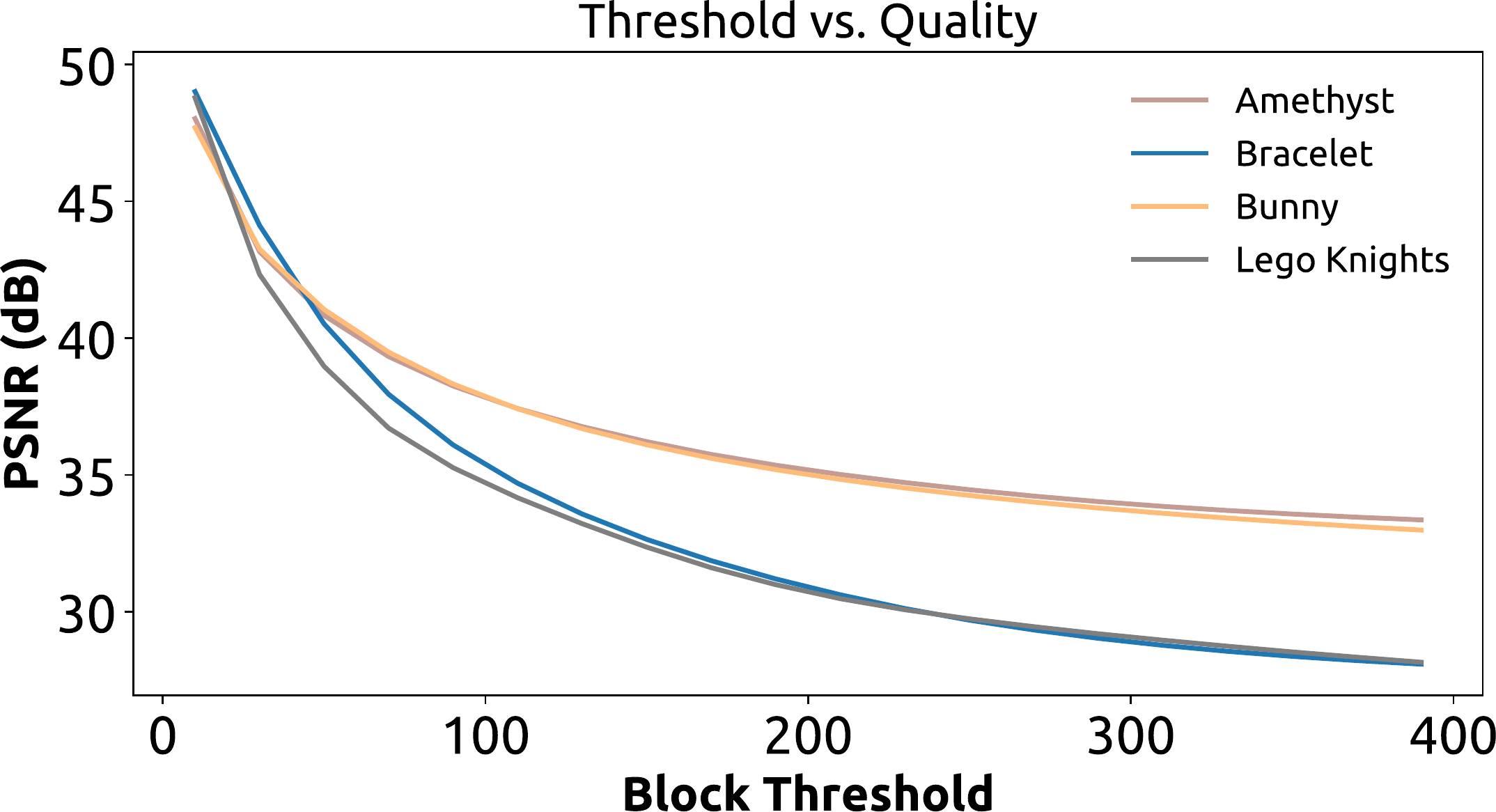}
\vspace*{-1em}
 \caption{The variation of the decompression quality (PSNR) with change in the \textit{block threshold} is plotted. The encoding parameters are set as: \textit{tree height} to 3, \textit{block size} to 2, \textit{quantization level} to no quantization and the \textit{block threshold} is varied. With an increase in the \textit{block threshold} the thresholding errors increase and the compression quality (PSNR) for all the LF datasets decreases.}
 \label{fig:thrsh_psnr}
\vspace*{-2em}
 \end{figure}

\begin{figure}[t!]
 \centering
 \includegraphics[width=\columnwidth, keepaspectratio=true]{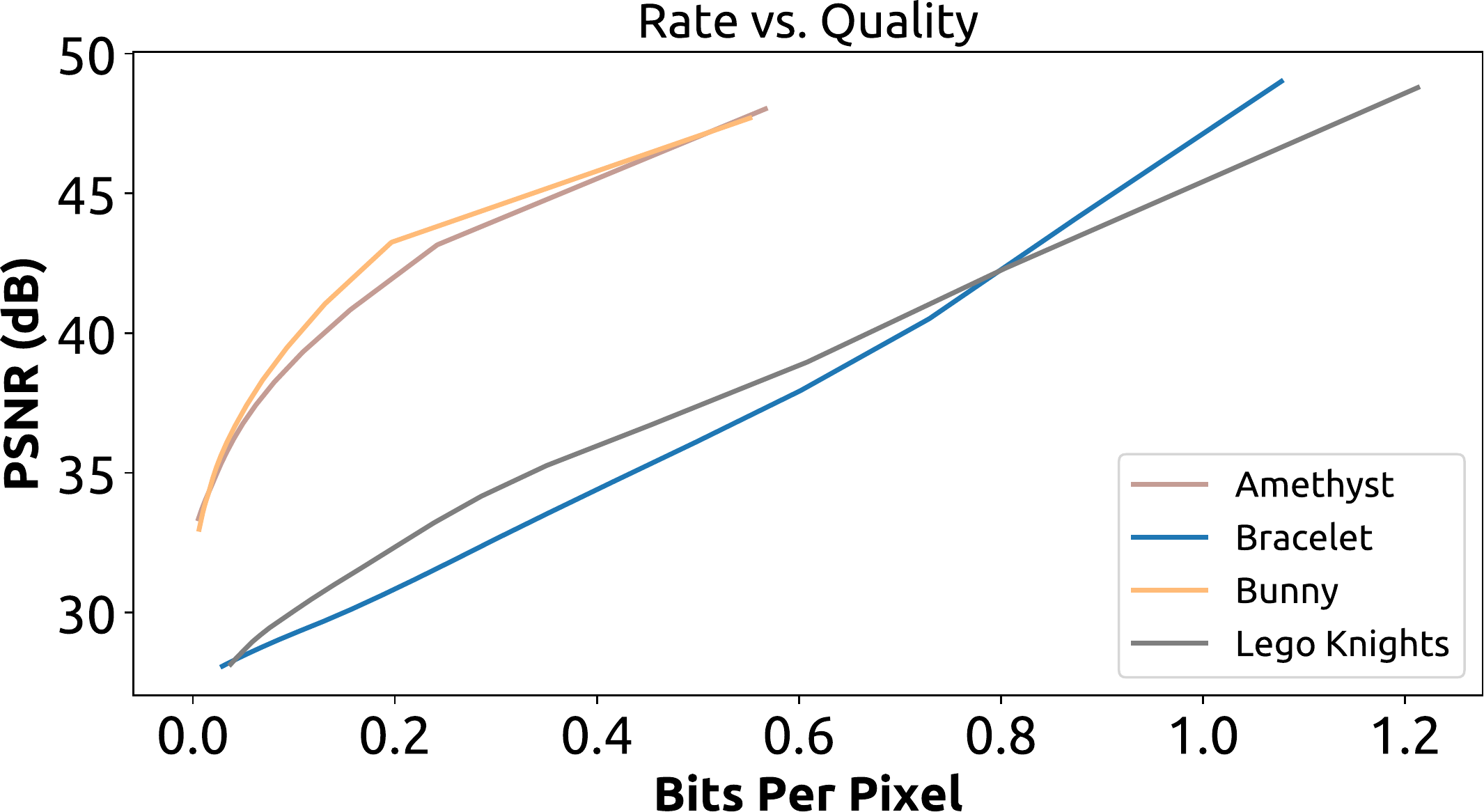}

 \caption{The variation of the decompression quality (PSNR) with change in compression rate (bpp) highlighted. The \textit{tree height} to 3 and \textit{block size} to 2 and vary the \textit{block threshold} and resulting bpp and PNSR values are plotted for the LF datasets.}
\vspace*{-1em}
 \label{fig:bpp_quality}
 \end{figure}

Figure~\ref{fig:thrsh_bpp} and Figure~\ref{fig:thrsh_psnr} show the outcome of varying the \textit{block threshold} on the bpp and PSNR, respectively. An increase in the \textit{block threshold} implies an increase in the threshold errors of the sparse residual blocks, resulting in a decrease of bpp and PSNR. The variation of the increase in the resulting quality with an increase in the bpp is shown in Figure~\ref{fig:bpp_quality}. The variation of the decompression quality with bpp is subject to the details in the LF under consideration. Figure~\ref{fig:zoomed_in} shows a zoomed in visual quality comparison of interesting regions of a few images from the LF dataset. The comparison shows that our compression method introduces no visible artifacts in the LF images.

\begin{table}[t!]
\begin{tabular}{|c|c|c|c|c|}
\hline
LF Dataset & Metric & \begin{tabular}[c]{@{}c@{}}Block\\ Size: 2\end{tabular} & \begin{tabular}[c]{@{}c@{}}Block\\ Size: 4\end{tabular} & \begin{tabular}[c]{@{}c@{}}Block\\ Size: 8\end{tabular} \\ \hline 
\multirow{2}{*}{Amethyst} & bpp & 0.109 & 0.34  & 1.12\\ \cline{2-5}
                           & PSNR & 40.82 & 45.97 & 49.56\\ \hline \hline
\multirow{2}{*}{Bunny} & bpp & 0.096 & 0.29 & 1.101\\ \cline{2-5}
                      & PSNR & 40.86 & 43.76 & 48.3\\ \hline \hline
\multirow{2}{*}{Bracelet } & bpp & 0.73  & 1.16 & 1.84 \\ \cline{2-5}
                           & PSNR & 40.51 & 48.45 & 52.85 \\ \hline \hline 
\multirow{2}{*}{Knight} & bpp & 0.498  & 0.855 & 1.44 \\ \cline{2-5}
                           & PSNR & 40.58 & 46.70 & 48.35 \\ \hline    
\end{tabular}
\caption{We highlight the variation in the resulting compression rates and qualities, as the  \textit{block size} changes. we set the \textit{block threshold} to $50$, the \textit{tree height} to 3, and the \textit{quantization level} to 2 bits for all the datasets. For a fixed \textit{block threshold}, the thresholding errors decrease when the \textit{block size} increases. As a result, we observe an increase in bpp and PSNR.}
\label{tab:blksize}
\vspace*{-3em}
\end{table}

\noindent {\bf Decoding time and frame rates:} We have implemented both GPU and CPU decoders to measure the decoding times. To compute a required final pixel value, each of the pixel values from all the channels (YCoCg26) are decoded independently and the corresponding RGB pixel value is computed. The average decode times to decode a block of pixels on an NVIDIA GTX-960 and an Intel Xeon 2.4GHz are: \begin{enumerate*}
\item Y-Channel: $2-3$ microseconds
\item Co-Channel: $1-2$ microseconds
\item Cg-Channel: $1-2$ microseconds.
\end{enumerate*}
We have implemented a parallelized GPU LF renderer that uses our decompression scheme to render new views. It takes $3-7$ milliseconds to generate a view with a resolution of $512 \times 512$ for a given new camera location, resulting in average frame rates of $200$ fps. While rendering a new view ($512 \times 512$), the average time taken by all the GPU threads are as follows: memory read-gather operations, about $4$ milliseconds; decoding operations, about $2$ milliseconds; ray-tracing and other computations, close to $1$ millisecond. For higher resolutions of $1024 \times 1024$, our renderer takes $8-13$ milliseconds per frame. The rendering is performed and the decoding times are measured on a file compressed using the encoding parameters: \textit{block size 4}, \textit{tree height 3}, and on the Lego Knights benchmark. It turns out that the total number of blocks that must be decoded grows linearly with the number of pixels. Our algorithm performs parallel decoding on the GPU using our compressed stream structure and the frame rate decreases at a sub-linear rate. 

{\small
\begin{table}[t!]
\begin{tabular}{|c|c|c|c|c|}
\hline
LF Dataset  & Metric & RLFC & \begin{tabular}[c]{@{}c@{}}Motion  \\ vectors\end{tabular} & \begin{tabular}[c]{@{}c@{}}Improvement  \\ factor\end{tabular}\\ \hline \hline
\multirow{2}{*}{Amethyst} & bpp & 0.109 & 0.197 & \cellcolor{green}1.80  \\ \cline{2-4}
                          & PSNR & 41.99 & 40.88 & \cellcolor{green} \\ \hline \hline
                           
\multirow{2}{*}{Bracelet} & bpp & 0.541 & 0.697 & \cellcolor{green}1.30 \\ \cline{2-4}
                          & PSNR & 41.31 & 42 & \cellcolor{green} \\ \hline \hline
                               
\multirow{2}{*}{Bunny} & bpp & 0.094  & 0.091 & \cellcolor{green}1.03 \\ \cline{2-4}
                      & PSNR & 40.85 & 41.6 & \cellcolor{green}\\ \hline \hline
                           
\multirow{2}{*}{Knights } & bpp & 0.62  & 0.35 & \cellcolor{red}1.71 \\ \cline{2-4}
                              & PSNR & 40.64 & 40.32 & \cellcolor{red}\\ \hline \hline
                           
\multirow{2}{*}{Tarot } & bpp & 2.2  & 1.2 & \cellcolor{red}1.83 \\ \cline{2-4}
                             & PSNR & 41.99 & 41.72 & \cellcolor{red}\\ \hline                            
\end{tabular}
\caption{We compare RLFC with motion compensation based compression scheme in terms of compression rates (bpp) for similar compression quality (PSNR). The last column indicates the improvement factor compared to RLFC. The cells are highlighted in green indicate cases where RLFC performs better or equal than the motion compensation scheme by the mentioned factor. The cells highlighted in red indicate the cases where motion compensation performs better than RLFC.}
\label{tab:compari}
\vspace*{-3em}
\end{table}
}

\noindent {\bf Comparison with random access LFI schemes:} Our hierarchical approach is orthogonal to the motion compensation schemes because we use totally different series of steps to exploit the redundancies among the LFI to achieve compression. RLFC offers several advantages compared to other methods: 
\begin{enumerate}
    \item The steps used in our compression scheme are simple and enable fast real-time encoding.
    \item The hierarchy computed in our method provides progressive streaming and decoding capabilities.
    \item BISE decoding used in our approach is currently supported on all major mobile platforms and thereby making it easy to implement our approach on mobile platforms. 
    \end{enumerate}

Our method offers at least 10X better compression ratios over the VQ method~\citep{LFLevoy96} and the 4D wavelet scheme~\citep{LFPeter01}. To compare our method with motion compensation schemes that provide random access we have implemented a motion compensation scheme that provides random access based on ~\citet{zhang2000compression}. The comparison of the resulting bit-rates for similar PSNR quality on different datasets is presented in Table~\ref{tab:compari}. For some LFI (Amethyst, Bracelet, and Bunny) RLFC achieves  similar or better compression rates in comparison with the motion compensation scheme. In other cases (Lego Knights, Tarot Cards), the motion compensation scheme achieves better compression rates compared to RLFC for similar quality. More comparisons similar to Table~\ref{tab:compari} are included in suppl. material, Section-4. 



On an LF dataset (Lego Bulldozer) from the Stanford LF archive with complex details in the scene and with a high resolution of $1536\times1152$ ~\citet{WelcomeLF} reports a compression ratio of 178:1 for PSNR of 45 dB. On the same LF dataset, our method achieves a compression of a ratio 60:1 for a PSNR of 42 dB. In comparison with RLFC motion compensation schemes provides better ($\sim2-3\times$) compression ratios on LFI with intricate details and large areas of high-frequency components. In other cases, RLFC provides better ($\sim2\times$) or similar compression ratios compared to the motion compensation scheme. In the case of LFI with intricate details and large areas of high frequency regions the SRVs (computed as difference of filtered RKV between alternate levels) are less sparse resulting in higher compression ratios. This can be noticed in the example visualization of the RLFC trees presented in suppl. material, Section-5.

\section{Conclusions, Limitations \& Future Work}
\label{sec:concl}

\textbf{Conclusion}: We present a new method (RFLC) that encodes LFI by constructing a hierarchy based on computing new sets of images (key views). Our method provides random access to the LF pixel values with one level of indirection and  supports parallel and progressive decompression. We have implemented our method on the two-plane parameterization of the light fields and highlight its performance. The average time to decode a block of pixels is $1-3$ microseconds per channel and can be used it for interactive rendering. Our method is simple, general, and also hardware friendly.

\textbf{Limitations}: Our approach has some limitations. The reconstruction quality of our method for a sparsely sampled light fields can be low. Our algorithm uses filtering of spatially close by LF samples to compute the representative key views. In a sparsely sampled light field, the nearby samples may not exhibit a high level of spatial correlation. Moreover, our method  doesn't provide fine grained control over of the reconstruction quality with respect to the encoding parameters. Therefore, a  change in the parameters can  affect the quality.  
Our current implementation of a GPU based LF renderer is unoptimized in memory access patterns of GPU threads and can be improved.

\textbf{Future Work}: We would like to extend and evaluate RLFC on other parameterizations including spherical~\citep{LFSphereihm1997} and unstructured LF~\citep{LFdavis2012unstructured}. The performance of our algorithms can be further improved using a dedicated hardware implementation. In the future, we would like to evaluate our method for very high-resolution LF datasets ($2160\times1200$ for HTC Vive and $1080\times1200$ for Oculus Rift) on commodity VR headsets and mobile devices. Currently, we use a uniform and localized clustering step. It will be useful to investigate better or global clustering schemes to improve the performance. We compute the representative key views at each level using weighted filtering, and it would be useful to explore other techniques based on motion vectors on a block level or image warping methods. Compared to motion compensation schemes our approach is complimentary and involves a different set of steps to exploit redundancies in LFI to achieve  good compression ratios. It would be useful to explore techniques to combine our hierarchical method with the motion compensation techniques to further improve the compression ratios~\citep{arxivHMLFC}. However, including motion compensation techniques into our scheme may add more overhead to the decoding algorithm and we need to evaluate the pros and cons carefully. The fast performance of RLFC makes it possible to integrate it with an interactive LF capturing and rendering system for a realtime LF capture and rendering system.

\section{Acknowledgements}
This research is supported in part by Intel. 


\end{document}